# Switched latent force models for reverse-engineering transcriptional regulation in gene expression data

Andrés F. López-Lopera, and Mauricio A. Álvarez

**Abstract**—To survive environmental conditions, cells transcribe their response activities into encoded mRNA sequences in order to produce certain amounts of protein concentrations. The external conditions are mapped into the cell through the activation of special proteins called transcription factors (TFs). Due to the difficult task to measure experimentally TF behaviours, and the challenges to capture their quick-time dynamics, different types of models based on differential equations have been proposed. However, those approaches usually incur in costly procedures, and they present problems to describe sudden changes in TF regulators. In this paper, we present a switched dynamical latent force model for reverse-engineering transcriptional regulation in gene expression data which allows the exact inference over latent TF activities driving some observed gene expressions through a linear differential equation. To deal with discontinuities in the dynamics, we introduce an approach that switches between different TF activities and different dynamical systems. This creates a versatile representation of transcription networks that can capture discrete changes and non-linearities We evaluate our model on both simulated data and real-data (e.g. microaerobic shift in E. coli, yeast respiration), concluding that our framework allows for the fitting of the expression data while being able to infer continuous-time TF profiles.

✦

## 1 INTRODUCTION

Transcriptional regulation is the biological process in living organisms in which the cell regulates the conversion of DNA to RNA (transcription) in response to external stimuli [1], [2]. It is well known that the source stimuli are mapped into the cell by special proteins called transcription factors (TFs). According to the needs of the cell, TF regulators may bind to specific genes to activate or inhibit their transcription in order to control mRNA activity [2], [3]. The information encoded within the mRNA sequence is translated to the ribosome, place in which it is synthesized into amino-acid chains to produce certain amounts of specific protein concentrations [1]. Self-regulation in cells is carried out by controlling the production of these concentrations, allowing cells to adapt to changes in environmental conditions. For this reason, the understanding of the transcriptional regulation in gene expressions is key for describing the biophysics, genetic and molecular basis, and it has been widely demanded in biomedical and bioengineering applications [1], [2], [4]. Drug design, bacterial transcription, and genetic engineering are some case studies in which transcriptional regulation contains important biological information [5]–[8].

Microarray technology are usually used to measure mRNA activities on a genome-wide scale. Techniques such as chromatin immunoprecipitation (ChIP) have largely unveiled the wiring of cellular transcriptional regulatory networks, identifying which mRNA are bound by which TF regulators [7]. More recently, high throughput sequencing methods have been developed to provide more precise measurements for gene expression quantification [9], [10]. However, while specific genes are relatively easy to measure experimentally, there are some challenges in measuring active TF concentration levels (or to evaluate their effect on genes) [7]. The comprehension of both gene expression activity and TF behaviour, together with the knowledge of key biological parameters, are needed to obtain a fully quantitative description of transcriptional processes. Although important, transcriptional regulation in gene expression data is far from being wholly understood [11].

Aiming to develop probabilistic approaches for modelling transcriptional processes, the statistical community has proposed several methods to infer TF activities [6]–[8], [11]–[17]. Current studies suggest that reverse-engineering approaches based on differential equations provide high accurate performances [7], [11], [15], [16]. In [12], a parametric approach based on a time-dependent linear ordinary differential equation (ODE) is introduced in order to describe gene activities. The approach presents two challenges. First, it is based on Markov Chain Monte Carlo (MCMC), requiring substantial computational resources to carry out the inference of TF concentrations [12]. Second, the approach limits the inference of TF regulators to discrete time-points where the data were collected. Later in [7], the authors show that Gaussian processes, inspired by the physical model from [12], provide a simple and computationally efficient non-parametric method for reverse-engineering of continuous-time TF profiles [7]. The aim from [7] is to infer the TF activity when there are no observations of its behaviour but using the data available from gene expressions. In particular, they introduce Gaussian

● *A. F. López-Lopera is with the Department of Mathematics and Industrial Engineering, Écoles des Mines de Saint-Étienne, Saint-Étienne, France, 42000. M. A. Álvarez is with the Department of Computer Science, University of Sheffield, Sheffield, UK S1 4DP*
*E-mail: andres-felipe.lopez@emse.fr, mauricio.alvarez@sheffield.ac.uk*

priors over the activities of latent TFs, and compute the corresponding Gaussian process of the mRNA profiles for which the mechanistic model is encoded into its covariance function. The hybrid framework from [7] greatly outperforms the approach from [12], making it an attractive model for more realistic regulatory networks. The proposal from [7] is extended later in [13] to infer latent chemical species in biochemical networks exhibiting also accurate results.

The approaches proposed in [7], [12]–[15] may be applied effectively when modelling dynamics of simple regulatory networks, where the gene expression data are driven by a finite set of TF regulators. However, in order to survive to environmental changes, cells continuously control the production of the proteins commonly exhibiting sudden changes in the TF activities. Sudden oxygen starvation in bacterium organisms is an example where the TF activity changes suddenly from an inactive state to an active state in order to activate or inhibit its transcription. Here, the absence of oxygen provokes that the organism moves from a nitric metabolism (without oxygen) to an aerobic (with oxygen) metabolism. This type of changes in the TF dynamics due to external stimuli demand proper models, representing interesting challenges to the computational biology community [8], [16], [18]. To the best of our knowledge, there are only two approaches in which quick time-varying behaviours of TF regulators are taken into account. The first approach is the switched-based model of continuous-time TF protein proposed in [8], where the TF regulator is stimulated by stress signals. In [8], the latent process in [7] is modelled as a Markovian stochastic process accounting for the transitions between two states. The model from [8] is later extended in [18] for learning combinatorial transcriptional dynamics from gene expression data. The second approach is the multi-switch model proposed in [16]. It consists of a piecewise-linear ODE model of mRNA dynamics, which can be fitted with a reversible jump MCMC sampler for estimating the gene-specific hyperparameters [16]. Approaches from [8] and [16] have shown satisfactory results to fitting the switching changes in TF behaviours, but computational overheads are significant due to complex inference algorithms.

Inspired by the idea of combining Gaussian processes with first order ordinary differential equations for modelling transcriptional regulation, in this paper we introduce a switched dynamical latent force model [19] that accounts for the quick time-varying behaviour of transcription factors in different transcriptional regulation processes. Latent force models [14], [15] are Gaussian processes with covariance functions derived from principled mechanistic models expressed as ordinary or partial differential equations. The Gaussian process model proposed by [7] can be seen as a particular case of a latent force model, one for which the covariance function is derived from a first order ordinary differential equation. Whereas a traditional latent force model assumes that the source underlying the mechanistic model is smooth, a switched dynamical latent force model [19] allows for discontinuities or non-smooth transitions for the latent input. In this paper, sharp transitions in a latent source can be either used to represent the on/off switching of different transcription factors or a transcription factor with sudden changes in its dynamics. Switched dynamical latent force models were introduced by [19] for segmenting motor skills for humanoid robotics applications. Motor skills are represented using second order ordinary differential equations.

TABLE 1
Mathematical notations

| Symbol | Definition |
|---|---|
| $a, A$ | Constants, hyperparameters |
| $\mathbf{x}, \mathbf{m}, \boldsymbol{\mu}$ | Column vectors |
| $\mathbf{X}, \mathbf{K}, \boldsymbol{\Sigma}$ | Matrices |
| $\mathbf{X}^\top, \mathbf{X}^{-1}$ | Transpose/Inverse of $\mathbf{X}$ |
| $y(t)$ | Function with entry $t$ |
| $y(t_1, t_2, \cdots, t_n)$ | Function with multiple entries $t_1, t_2, \cdots, t_n$ |
| $\{y_d(t)\}_{d=1}^D$ | A set of functions $y_d(t)$ where $d = 1, \cdots, D$ |
| $\operatorname{erf}\{x\}$ | Error function evaluated at $x$ |
| $p(x)$ | Marginal density distribution of $x$ |
| $p(x\|\theta)$ | Conditional distribution of $x$ given $\theta$ |
| $\mathcal{GP}(\mathbf{m}, \mathbf{K})$ | Gaussian process with mean $\mathbf{m}$ and covariance $\mathbf{K}$ |
| $\mathcal{N}(\mathbf{m}, \mathbf{K})$ | Gaussian distribution |
| $\mathbf{K}_{\mathbf{x},\mathbf{y}}$ | Covariance matrix with entries $k_{x,y}(t, t')$ |
| $k_{x,y}(t, t')$ | Covariance function between $x(t)$ and $y(t')$ |

The switched dynamical latent force model constrains the gene activity (output) at each switching time to be the same, allowing for the gene expressions to remain continuous and piecewise differentiable. It also provides an exact and computationally efficient framework for inference over the latent TF profiles.

We test our model under different artificial toy data and real-world biological problems (e.g. microaerobic shift in E. coli, and the control of ribosomal production in yeast metabolic cycle). For the real-data examples, we compare the results of our approach with respect to the ones proposed in [8], [18], due to the availability of their codes to reproduce the examples.

This paper is organised as follows. The background is described in section 2. We introduce the standard latent force model (LFM) proposed in [14]. In section 3, we derive the switched dynamical LFM when modelling reverse-engineering transcriptional regulation for quick time-varying TF regulators. The procedure followed for the experimental results is given in section 4. In section 5, we show and discuss the results obtained employing our framework on both artificial data and real-world biological data. Section 6 shows the conclusions, as well as the potential future work. Finally, in the appendices, we describe in detail the computation of some necessary expressions required for the formulation of the model.

## 2 BACKGROUND

In this section, we focus on the description of reverse-engineering transcriptional regulation, and review the standard latent force model (LFMs) using Gaussian processes (GPs) for reverse-engineering transcriptional regulation. Table 1 shows the mathematical notations employed in this paper.

## 2.1 Reverse-engineering for transcriptional regulation in gene expression data

In order to develop probabilistic reverse-engineering approaches for transcription networks, several models have been proposed to infer activity levels of proteins from time-series measurements of the targets' expression levels [7], [8], [11]–[18], [20]–[22]. Recent studies have shown that there is a simplification to model transcriptional regulation processes in which the gene expression data involve the use of differential equations [7], [13], [16], [18]. To describe the reverse-engineering transcriptional regulation in gene expression data, different mechanistic models have been employed: from linear ordinary differential equations (ODE) [7], [8], [11], [13]–[18], [21], to non-linear partial differential equations (PDE) [15], [20]–[22]. However, since the early days of molecular biology, the mechanistic model based on the first-order ODE from [1] has shown accurate results with high-resolution dynamics experiments [1], [2].

Later in [7], [15], the physical system from [1] is extended for multiple gene activities in which the contribution of external stimuli are included. According to [7], [15], a set of $D$ gene expressions $\{y_d(t)\}_{d=1}^D$ (outputs), which are driven by a set of $R$ TF regulators $\{u_r(t)\}_{r=1}^R$ (driven-forces), can be modelled using the following set of $D$ coupled first-order ODEs

$$\frac{dy_d(t)}{dt} + \gamma_d y_d(t) = B_d + \sum_{r=1}^R S_{r,d} u_r(t), \quad (1)$$

where $B_d$ and $\gamma_d$ correspond to the basal transcription and the decay rate of the $d$-th gene, respectively. The term $S_{r,d}$ represents the sensitivity of gene $d$ with respect to the protein $u_r(t)$. We note that the Equation (1) assumes that the transcript is degraded proportionally to its concentration, with a degradation rate $\gamma_d$. The production term $B_d + \sum_{r=1}^R S_{r,d} u_r(t)$ comprises the basal transcription rate $B_d$, which may vary proportionally according to the protein activities $\{u_r(t)\}_{r=1}^R$ [12]. The mechanistic model from Equation (1) can be seen as an overly simplified system, but it is amongst the methods with more accurate performance for modelling biological and bacterial transcription networks [7], [8], [11]–[15].

## 2.2 Latent force models using Gaussian processes

Latent force models (LFMs) using Gaussian processes (GPs) have been introduced first in [14], [15] as a hybrid model to combine data-driven modelling with physical models. For standard LFMs, the mechanistic model is encoded into the mean and the covariance function of the GP obtained for the outputs, and between the outputs and latent inputs [15], [21]. Initially, LFMs have been introduced to model mainly two demanding real-world applications based on ODE: human motion capture data, and reverse-engineering transcriptional regulation [14]. However, LFM may also be applied to spatio-temporal problems described by partial differential equations (PDE) [15], [21], and in other types of research fields such as neuroscience [23].

For reverse-engineering transcriptional regulation in gene expression, using the mechanistic model from Equation (1), an LFM assumes that the TF regulators $\{u_r(t)\}_{r=1}^R$ are unknown functions (latent forces) due to the fact that the TF behaviours are commonly difficult to measure experimentally. For this reason, some prior assumptions are required over each latent function $u_r(t)$. If we assume that $u_r(t)$ follows a zero-mean GP prior with covariance function $k_{u_r,u_r}(t,t')$,

$$u_r(t) \sim \mathcal{GP}(0, k_{u_r,u_r}(t,t')), \quad (2)$$

then, due to linearity of the Equation (1), the output $y_d(t)$ follows a GP with mean function $m_d(t)$ and covariance function $k_{y_d,y_d}(t,t')$. This is

$$y_d(t) \sim \mathcal{GP}(m_d(t), k_{y_d,y_d}(t,t')), \quad (3)$$

where $m_d(t)$ and $k_{y_d,y_d}(t,t')$ depend on the mechanistic model, and they contain the biological properties from Equation (1). Furthermore, the cross covariance function $k_{y_d,u_r}(t,t')$ between $y_d(t)$ and $u_r(t)$ can also be computed for inference purposes [15], [24]. The resulting joint GP from the LFM approach can be written as a joint multivariate Gaussian distribution for a finite number of time points, obtaining

$$\begin{bmatrix} \mathbf{u} \\ \mathbf{y} \end{bmatrix} \sim \mathcal{N}\left(\begin{bmatrix} \mathbf{0} \\ \mathbf{m} \end{bmatrix}, \begin{bmatrix} \mathbf{K}_{\mathbf{u},\mathbf{u}} & \mathbf{K}_{\mathbf{y},\mathbf{u}}^\top \\ \mathbf{K}_{\mathbf{y},\mathbf{u}} & \mathbf{K}_{\mathbf{y},\mathbf{y}} \end{bmatrix}\right).$$

The latent vector is given by $\mathbf{u} = [\mathbf{u}_1^\top, \cdots, \mathbf{u}_R^\top]^\top$, where $\mathbf{u}_r \in \mathbb{R}^{N \times 1}$ contains $N$ evaluations of the function $u_r(t)$ at particular points $t = \{t_n\}_{n=1}^N$. The vectors $\mathbf{y} = [\mathbf{y}_1^\top, \cdots, \mathbf{y}_D^\top]^\top$, and $\mathbf{m} = [\mathbf{m}_1^\top, \cdots, \mathbf{m}_D^\top]^\top$, are defined in a similar way to $\mathbf{u}$: the vectors $\mathbf{y}_d \in \mathbb{R}^{N \times 1}$, and $\mathbf{m}_d \in \mathbb{R}^{N \times 1}$, contain the evaluations of the output function $y_d(t)$ and the mean function $m_d(t)$, respectively. The terms $\mathbf{K}_{\mathbf{u},\mathbf{u}}$, $\mathbf{K}_{\mathbf{y},\mathbf{u}}$, and $\mathbf{K}_{\mathbf{y},\mathbf{y}}$ are covariance matrices with entries given by the functions $k_{u_r,u_{r'}}(t,t')$, $k_{y_d,u_{r'}}(t,t')$, and $k_{y_d,y_{d'}}(t,t')$, respectively.

Additionally, we are interested in the likelihood $p(\mathbf{y}|\boldsymbol{\theta})$, used for estimating biological parameters $\boldsymbol{\theta} = \{\gamma_d, B_d, S_d\}_{d=1}^D$. We can also use the likelihood to estimate any additional hyperparameters associated with the GP prior over the functions $u_r(t)$. By using the Bayes' theorem and the Gaussian properties, it is possible to infer the TF activity through the computation of the posterior distribution $p(\mathbf{u}|\mathbf{y})$ [25], [26].

## 2.3 LFMs for reverse-engineering transcriptional regulation in gene expression data

Reverse-engineering transcriptional regulation in gene expression data can be formulated as an LFM in which the mechanistic model is governed by the model from Equation (1). As we described in subsection 2.2, we need to compute the functions $m_d(t)$ and $k_{y_d,y_{d'}}(t,t')$ to build the GP over the outputs $\{y_d(t)\}_{d=1}^D$. First, we need to compute the solution for Equation (1), which is given by [7], [15]

$$y_d(t) = [1 - c_d(t)]\frac{B_d}{\gamma_d} + c_d(t)y_d(0) + \sum_{r=1}^R f_{r,d}(t,u_r), \quad (4)$$

where $y_d(0)$ is the initial gene expression abundance level, and terms $c_d(t)$ and $f_{r,d}(t,u_r)$ are defined as

$$c_d(t) = \exp\{-\gamma_d t\},$$

$$f_{r,d}(t,u_r) = S_{r,d} c_d(t) \int_0^t u_r(\tau) \exp\{\gamma_d \tau\} d\tau.$$

We note that the term $f_{r,d}(t,u)$ has an implicit dependence on the force $u_r(t)$. The uncertainty on the output $y_d(t)$ comes from the uncertainty over $\{u_r(t)\}_{r=1}^R$ and the uncertainty over the initial condition $y_d(0)$.

We make the following assumptions in order to build the GP over the outputs $\{y_d(t)\}_{d=1}^D$. First, we assume that the initial conditions, $\mathbf{y}_{IC} = [y_1(0), y_2(0), \cdots, y_D(0)]^\top$, follow a zero-mean Gaussian distribution with covariance $K_{IC}$. We also assume that the initial conditions are independent of the latent forces $\{u_r(t)\}_{r=1}^R$, plus we place an independent GP prior over each latent force $u_r(t)$. It then follows that the mean function $m_d(t)$ is given by

$$m_d(t) = [1 - c_d(t)]\frac{B_d}{\gamma_d}. \quad (5)$$

The covariance function $k_{y_d,y_{d'}}(t,t')$ between any two output functions, $d$ and $d'$, at any two times, $t$ and $t'$, is given by

$$k_{y_d,y_{d'}}(t,t') = c_d(t)c_{d'}(t')\sigma_{y_d,y_{d'}} + \sum_{r=1}^R k_{f_{r,d},f_{r,d'}}(t,t'), \quad (6)$$

where $\sigma_{y_d,y_{d'}}$ are entries of covariance matrix $K_{IC}$, and the covariance $k_{f_{r,d},f_{r,d'}}(t,t')$ is defined as

$$k_{f_{r,d},f_{r,d'}}(t,t') = S_{r,d}S_{r,d'}c_d(t)c_{d'}(t') \times \int_0^t \exp\{\gamma_d \tau\} \int_0^{t'} \exp\{\gamma_{d'}\tau'\} k_{u_r,u_r}(\tau,\tau')d\tau' d\tau.$$

We see that the covariance $k_{f_{r,d},f_{r,d'}}(t,t')$ depends on the covariance function between the latent forces $k_{u_r,u_r}(t,t')$. If we assume that each kernel $k_{u_r,u_r}(t,t')$ follows a Squared Exponential (SE) kernel function given by

$$k_{u_r,u_r}(t,t') = \exp\left\{-\frac{(t-t')^2}{\ell_r^2}\right\}, \quad (7)$$

where $\ell_r$ is known as the length-scale parameter related to $u_r(t)$, then $k_{f_{r,d},f_{r,d'}}(t,t')$ can be computed analytically [7]

$$k_{f_{r,d},f_{r,d'}}(t,t') = \frac{S_{r,d}S_{r,d'}\ell_r\sqrt{\pi}}{2}[\hat{h}(\gamma_{d'},\gamma_d,t,t') + \hat{h}(\gamma_d,\gamma_{d'},t',t)], \quad (8)$$

where the function $\hat{h}(\gamma_{d'},\gamma_d,t,t')$ is given by

$$\hat{h}(\gamma_{d'},\gamma_d,t,t') = \frac{1}{\gamma_d + \gamma_{d'}}\left[\Upsilon(\gamma_{d'},t',t) - \exp\{-\gamma_d t\}\Upsilon(\gamma_{d'},t',0)\right],$$

with

$$\Upsilon(\gamma_{d'},t',t) = \exp\{\nu_{r,d'}^2\}\exp\{-\gamma_{d'}(t'-t)\} \times \left[\operatorname{erf}\left\{\frac{t'-t}{\ell_r} - \nu_{r,d'}\right\} + \operatorname{erf}\left\{\frac{t}{\ell_r} + \nu_{r,d'}\right\}\right],$$

with $\nu_{r,d'} = \ell_r \gamma_{d'}/2$, and $\operatorname{erf}\{\cdot\}$ is the *error function*.

We also need to compute the cross-covariance between the $d$-th output function and the latent force at any two times, $t$ and $t'$, to complete the joint multivariate Gaussian distribution from subsection 2.2. Such covariance follows as

$$k_{y_d,u_r}(t,t') = S_{r,d}c_d(t)\int_0^t \exp\{\gamma_d \tau\} k_{u_r,u_r}(\tau,t')d\tau.$$

Expression for $k_{y_d,u_r}(t,t')$ can also be computed analytically, and it is given by

$$k_{y_d,u_r}(t,t') = \frac{S_{r,d}\ell_r\sqrt{\pi}}{2}\Upsilon(\gamma_d,t,t'). \quad (9)$$

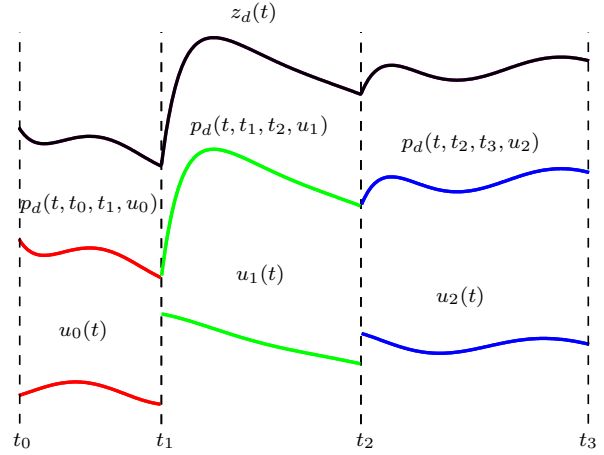

Fig. 1. Cartoon representation of output $z_d(t)$ switching its behaviour between points $t_0$, $t_1$, $t_2$ and $t_3$.

Then, using the covariances functions from Equations (7), (8), and (9), the posterior distribution $p(\mathbf{u}|\mathbf{y})$ (i.e. full inference of protein profiles given the gene expression data) can be computed according to [7], [24].

Apart from the advantages of using GP-based approaches (e.g. inference of continuous responses, handling of uncertainties) [24], the posterior covariance given by the LFM framework depends on the biological principles from Equation (1). In addition, as previous studies suggest, employing physically inspired covariances provide more accurate predictions even in regions where there is no available gene expression data [15]. More experimental and theoretical properties about using LFM for reversing-engineering transcription regulation are discussed in [7], [13], [21].

## 3 SWITCHED LFM FOR REVERSE-ENGINEERING TRANSCRIPTIONAL REGULATION

Exploiting the advantages of using GPs with physically inspired covariances for reverse-engineering transcriptional regulation, we now introduce a new LFM that allow the on and off switching of different latent forces in different time segments. This makes possible to express changes in the output of a dynamical system due to potentially non-continuous changes in the TF activity. We assume that the gene profiles in each segment are driven by $R$ independent TF regulators. We constrain the gene activities at each switching time to be the same, so that they remain continuous. Figure 1 shows a cartoon representation of gene expression $z_d(t)$ switching its behaviour in three non-overlapping segments, between points $t_0$, $t_1$, $t_2$ and $t_3$ when only one TF protein is driving the process in each segment ($R = 1$). That means that in each interval $(t_{q-1}, t_q)$, only the latent force $u_{q-1}(t)$ is active.

### 3.1 Definition of the model

First, we assume that the input space is divided in a series of non-overlapping intervals $[t_{q-1}, t_q]_{q=1}^Q$. During each interval, $R$ independent TF proteins $\{u_{r,q-1}(t)\}_{r=1}^R$ out of $R \times Q$ forces are active. The forces $\{u_{r,q-1}(t)\}_{r=1}^R$ are activated after time $t_{q-1}$ (switched on) and deactivated (switched off) after time $t_q$ [19]. We use the reverse-engineering model from Equation (1) to describe the contribution of outputs due to the sequential activation of forces $\{u_{r,q-1}(t)\}_{r=1,q=1}^{R,Q}$. For simplicity, we remove the mean of the process for further analysis. However, the mean function from the Equation (5) could easily be included in the GP regression formalism.

A particular output $z_d(t)$ at a particular time instant $t$, in the interval $(t_{q-1}, t_q)$, is expressed as

$$z_d(t, t_{q-1}, t_q) = p_d(t, t_{q-1}, t_q, u_{q-1}), \quad \text{for} \quad 1 \leq d \leq D,$$

where $p_d(t, t_{q-1}, t_q, u_{r,q-1})$ uses the model from Equation (4), and is equal to

$$p_d(t, t_{q-1}, t_q, u_{r,q-1}) = c_d(t - t_{q-1}) z_d(t_{q-1}) \\ + \sum_{r=1}^{R} f_{r,d}(t, t_{q-1}, t_q, u_{r,q-1}), \quad (10)$$

with

$$f_{r,d}(t, t_{q-1}, t_q, u_{r,q-1}) \\ = S_{r,d}^{(q-1)} c_d(t) \int_{t_{q-1}}^{t_q} u_{r,q-1}(\tau) \exp\{\gamma_d \tau\} d\tau. \quad (11)$$

Note that the sensitivity parameters $S_{r,d}^{(q-1)}$ contain one additional index compared to the standard LFM from subsection 2.3. This is because we need to define a set of sensitivity parameters $\{S_{r,d}\}_{r=1,d=1}^{R,D}$ for each interval when $q = 1, 2, \cdots, Q$. In this sense, the terms $S_{r,d}^{(q-1)}$ represent the sensitivity of gene $d$ respect to protein $u_{r,q-1}(t)$ at the interval $q$. The expression from Equation (10) is assumed to be valid for describing the output only inside the interval $(t_{q-1}, t_q)$. Note that the operator $f_{r,d}(t, t_{q-1}, t_q, u_{r,q-1})$ from Equation (11) is a function of four arguments: the first argument, $t$, refers to the independent variable; the second argument $t_{q-1}$ and the third argument $t_q$ specify the lower and upper limits of the time interval to be analysed (respectively); and the last argument, $u_{r,q-1}$, specifies the $r$-th latent force that is acting within the interval $q$.

Given the parameters $\boldsymbol{\theta} = \{\gamma_d, S_{r,d}^{(q)}, \ell_{r,q}\}_{d=1,r=1,q=0'}^{D,R,Q-1}$ the uncertainty in outputs is induced by the prior over the initial conditions $z_d(t_{q-1})$ for all values of $t_{q-1}$, and the latent forces $\{u_{r,q-1}(t)\}_{r=1}^{R}$ that are active during $(t_{q-1}, t_q)$. We place an independent GP prior over each latent force $u_{r,q-1}(t)$. For initial conditions $z_d(t_{q-1})$, we assume that they are also hyperparameters to be estimated or random variables with uncertainty governed by independent Gaussian distributions with covariance matrices $K_{IC}^q$. We also consider that the outputs should be continuous at the switching points. Therefore the uncertainty about initial conditions for interval $q$, $z_d(t_{q-1})$, is proscribed by the GP that describes the outputs $z_d(t)$ in the previous interval $q-1$. The random variables $z_d(t_{q-1})$ are then Gaussian-distributed with mean values given by $p_d(t_{q-1}, t_{q-2}, t_{q-1}, u_{q-2})$ and covariances $k_{z_d, z_{d'}}(t_{q-1}, t_{q'-1})$.

### 3.2 Covariance for the outputs

For continuous output signals, we must take into account the constrains at each switching time instant. Such constrains cause initial values for each interval to be dependent on final conditions for the previous interval and induce correlations across the intervals [19]. We are interested in the computation of covariance functions for the outputs, and between the outputs and the latent forces, as we described for the SIM framework in section 2.3. We need to compute the covariance $\text{cov}\{z_d(t, t_{q-1}, t_q), z_{d'}(t', t_{q'-1}, t_{q'})\}$ for $z_d(t, t_{q-1}, t_q)$ in time interval $(t_{q-1}, t_q)$, and $z_{d'}(t', t_{q'-1}, t_{q'})$ in time interval $(t_{q'-1}, t_{q'})$. For this reason, we have to analyse three regimes: $q > q'$, $q = q'$ and $q < q'$. We compute the first two cases, $q > q'$ and $q = q'$. The solution for $q < q'$ is given by $q > q'$, when the roles between $q$ and $q'$ are interchanged.

#### 3.2.1 Covariance for the interval $(t_{q-1}, t_q)$

For the covariance $k_{z_d, z_{d'}}(t, t') = \text{cov}\{z_d(t), z_{d'}(t')\}$ in the interval $(t_{q-1}, t_q)$, we obtain an expression similar to the one obtained in Equation (6) for the standard LFM,

$$k_{z_d, z_{d'}}(t, t') = c_d(t - t_{q-1}) c_{d'}(t' - t_{q-1}) k_{z_d, z_{d'}}(t_{q-1}, t_{q-1}) \\ + \sum_{r=1}^{R} k_{f_{r,d}, f_{r,d'}}^{(q)}(t, t'), \quad (12)$$

where the covariance $k_{z_d, z_{d'}}(t_{q-1}, t_{q-1})$ is given by the covariance between $p_d(t_{q-1}, t_{q-2}, t_{q-1}, u_{q-2})$ and $p_{d'}(t_{q-1}, t_{q-2}, t_{q-1}, u_{q-2})$ (considering outputs should be continuous across switching points), and the covariance $k_{f_{r,d}, f_{r,d'}}^{(q)}(t, t')$ is given by $\text{cov}\{f_{r,d}(t, t_{q-1}, t_q, u_{r,q-1}), f_{r,d'}(t', t_{q-1}, t_q, u_{r,q-1})\}$ which can be computed using Equation (8).

#### 3.2.2 Covariance for the interval $(t_{q-1}, t_q)$ and $(t_{q'-1}, t_{q'})$

When $q > q'$, we have to take into account the correlation between the initial condition $z_d(t_{q-1})$ and the latent forces $\{u_{r,q'-1}(t')\}_{r=1}^{R}$. This correlation appears due to the contribution of $\{u_{r,q'-1}(t')\}_{r=1}^{R}$ to generate the initial conditions $z_d(t_{q-1})$. We can then rewrite $k_{z_d, z_{d'}}(t, t')$ as

$$k_{z_d, z_{d'}}(t, t') = c_d(t - t_{q-1}) c_{d'}(t' - t_{q'-1}) k_{z_d, z_{d'}}(t_{q-1}, t_{q'-1}) \\ + c_d(t - t_{q-1}) \sum_{r=1}^{R} \text{cov}\{z_d(t_{q-1}), f_{r,d'}(t', t_{q'-1}, t_{q'}, u_{r,q'-1})\}. \quad (13)$$

For $q = q' + n$, we obtain the following recursive equation for the covariance between the initial condition $z_d(t_{q-1})$ and $f_{r,d'}(t', t_{q'-1}, t_{q'}, u_{r,q'-1})$

$$\text{cov}\{z_d(t_{q-1}), f_{r,d'}(t', t_{q'-1}, t_{q'}, u_{r,q'-1})\} \\ = \left[\prod_{i=1}^{n} c_d(t_{q-i} - t_{q-i-1})\right] k_{f_{r,d}, f_{r,d'}}^{(q'-1)}(t_{q-n}, t'),$$

where the term $k_{f_{r,d}, f_{r,d'}}^{(q'-1)}(t_{q-n}, t')$ is given by the covariance between $f_{r,d}(t_{q-n}, t_{q-n-1}, t_{q-n}, u_{r,q-n-1})$ and $f_{r,d'}(t', t_{q'-1}, t_{q'}, u_{r,q'-1})$, and it can be computed using Equation (8). In the appendix A, we show in detail the computation for the covariance of the outputs when only one TF ($R = 1$) is driven the output in each interval. The generalization of the model is straightforward for the case in which $R$ TF regulators are acting in each interval (assuming they are independent).

For the estimation of TF activities, we also need to compute the cross-covariances between the outputs $z_d(t, t_{q-1}, t_q)$ and the latent forces $u_{r,q'-1}(t)$. The computation of those covariances follows a similar procedure than the one employed for the covariance between the outputs, and its expression is given in the appendix B.

## 4 PROCEDURE

In this section, we describe additional details of the hyperparameters estimation, computational implementation, and biological datasets employed in this paper.

### 4.1 Hyperparameter estimation of covariance functions

Given the number of outputs $D$, the number of intervals $Q$, and the number of TF proteins $R$ acting in each interval, we estimate the parameters $\boldsymbol{\theta} = \{\gamma_d, S_{r,d}^{(q)}, \ell_{r,q}, t_{q+1}\}_{d=1,r=1,q=0}^{D,R,Q-1}$ by maximizing the marginal-likelihood of the joint GP $\{z_d^q(t)\}_{d=1}^D$ using gradient-descent methods [19], [24], [27]. Given a set of input discrete-time points, $\mathbf{t} = \{t_n\}_{n=1}^N$, the marginal-likelihood is given as $p(\mathbf{z}|\boldsymbol{\theta}) = \mathcal{N}(\mathbf{z}|\mathbf{m}, \mathbf{K}_{\mathbf{z},\mathbf{z}} + \boldsymbol{\Sigma})$, where $\mathbf{z} = [\mathbf{z}_1^\top, \cdots, \mathbf{z}_D^\top]^\top$, with $\mathbf{z}_d = [z_d(t_1), \cdots, z_d(t_N)]^\top$; $\mathbf{m} = [\mathbf{m}_1^\top, \cdots, \mathbf{m}_D^\top]^\top$, with entries $\mathbf{m}_d = [m_d(t_1), \cdots, m_d(t_N)]^\top$ given by the mean behaviour given from Equation (5); and $\mathbf{K}_{\mathbf{z},\mathbf{z}}$ is a $D \times D$ block-partitioned matrix with blocks $\mathbf{K}_{\mathbf{z}_d, \mathbf{z}_{d'}}$. The entries in each of these blocks are evaluated using $k_{z_d, z_{d'}}(t, t')$. Furthermore, $k_{z_d, z_{d'}}(t, t')$ is computed as we showed in subsection 3.2, according to the particular regimes of $q$ and $q'$ (see Equations (12) and (13)). Note that the computation of $m_d(t)$ and $k_{z_d, z_{d'}}(t, t')$ depend on the dynamic system, the mean and the covariance functions from the latent forces of $u_r(t)$, but they do not depend directly on $u_r(t)$. Finally, the covariance matrix $\boldsymbol{\Sigma}$ is included in order to take into account additive noises with variances $\sigma_d^2$, i.e. $\boldsymbol{\Sigma}$ is an identity block matrix, where the entries of the blocks is given by $\sigma_d^2 \delta_{d,d'}$.

### 4.2 Computational implementation

The multi-output switched dynamical latent force model for reverse-engineering transcriptional regulation in gene expression data was developed in MATLAB®, and the codes used in this paper are available on Github: https://github.com/anfelopera/SDLFM_ReverseEngineering. It is based on the GPmat Toolbox.[1] For the implementation of the codes, we work with the inverse of the length-scale defined as $\ell_{r,q} = (2/\hat{\ell}_{r,q})^{1/2}$. Algorithm 1 shows the pseudo-code for the implementation of the switched latent force model for reverse-engineering transcriptional regulation in gene expression data.

Results presented here that use the models from [8], [18], were reproduced using their corresponding source codes.

---

1. The GPmat toolbox provides several MATLAB® implementations of Gaussian processes based models. It is available in: https://github.com/SheffieldML/GPmat

---

**Algorithm 1** Switched latent force model for reverse-engineering

1: **procedure** PREDICTION OF TF PROFILES USING THE GENES PROFILES
2: Input: number of outputs $D$, number of intervals $Q$, number of latent forces $R$, genes profiles $\{z_d(t)\}_{d=1}^D$, initial set of hyperparameters $\boldsymbol{\theta} = \{\gamma_d, S_{r,d}^{(q)}, \ell_{r,q}, t_{q+1}\}_{d=1,r=1,q=0}^{D,R,Q-1}$
3: Compute the covariances matrices $\mathbf{K}_{\mathbf{z},\mathbf{z}}$, $\mathbf{K}_{\mathbf{u},\mathbf{u}}$, and $\mathbf{K}_{\mathbf{z},\mathbf{u}}$ according to sections 2 and 3.
4: Estimate the hyperparameters $\hat{\boldsymbol{\theta}}$ by maximizing the marginal-likelihood of the joint GP $\{z_d^q(t)\}_{d=1}^D$, i.e. $\hat{\boldsymbol{\theta}} = \arg\max_{\boldsymbol{\theta}}\{p(\mathbf{z}|\boldsymbol{\theta})\}$ (see subsection 4.1).
5: Using the partitioned Gaussian properties [24], compute the posterior of $p(\mathbf{u}|\mathbf{z}, \hat{\boldsymbol{\theta}})$.

### 4.3 Biological datasets

#### 4.3.1 Microaerobic shift in E. coli

The bacterium *Escherichia coli* (E. coli) has been extensively studied due to its quick adaptation to sudden changes in its environment [28]. This allows researchers to make routinely experiments over its robust organism without damaging the cells. For example, the sudden oxygen starvation in the E. coli environment provokes the bacterium organism changes from a nitric (absence of oxygen) metabolism to a much more energetically favourable aerobic metabolism (presence of oxygen) [8]. The E. coli dataset employed in [8], contains microarray data of five different gene expression levels (ompW, yjiD, hypB, moaA, and aspA), which are regulated by a TF protein known as FNR (Fumarate and Nitrate reductase Regulatory) [28]. Different microarray measurements represent changes in the concentrations of mRNA relative to the initial condition. There are measurements at five-time instants: 0, 5, 10, 15 and 60 min.

#### 4.3.2 Control of ribosomal production in yeast metabolism

This dataset describes the dynamics of control of ribosomal protein production in the metabolism of a specific type of eukaryotic microorganism. The phenomenon is known as yeast respiration [18]. The metabolic cycle was assayed using microarrays by Tu et al. in [29]. The dataset contains the measurements of 3178 gene expression levels, measured in 36-time points sampled at 25 min intervals through three cycles of yeast respiration. In each cycle, the organism is forced to a starvation of oxygen, followed by a constant supply of glucose for a period of time. Experimental studies of yeast respiration have shown that there are two important transcriptional regulators for controlling the production of ribosomal proteins, FHL1 (Four and a Half LIM domains 1) and RAP1 (RAs-related Protein 1) [18], [30], [31]. According to the ChIP-on-chip data,[2] there are ten genes that depend on both transcription proteins. First, the FHL1 protein regulates solely two genes: YLR030W and TKL2. Second, RAP1 protein regulates other two different genes: YOR359W and PFK27. The remaining five genes (RPL9A, RPL13A, RPL17B, RPL30 and RPS16B) are jointly regulated by FHL1 and RAP1, although the precise nature of the control is not known.

---

2. ChIP-on-chip is a technology that combines chromatin immunoprecipitation with DNA microarray. ChIP-on-chip is used to investigate interactions between proteins and DNA in vivo.

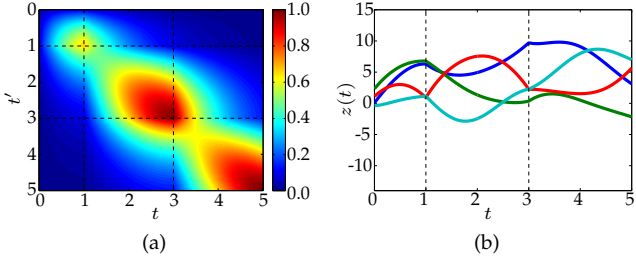

Fig. 2. Toy experiment 1. (a) covariance matrix between the outputs of the model with entries given by $k_{z,z}(t,t')$; with hyperparameters $D=1$, $Q=3$ and $R=1$; and switching points at $t_1=1$ and $t_2=3$. (b) samples generated from the zero-mean GP, $y(t) \sim \mathcal{GP}(0, k_{z,z}(t,t'))$.

## 5 RESULTS AND DISCUSSION

We propose several artificial experiments in order to highlight some properties of our approach that we cannot discuss in detail with the real study cases. Next, we evaluate the performance of our approach for reverse-engineering transcriptional regulation for the expression networks described in subsection 4.3. We compare our results with respect to the ones obtained by using the approaches proposed in [8], [18].

### 5.1 Toy examples

For the toy examples, we generate samples from a zero-mean GP with a covariance function explained in section 3. We implement several examples to evaluate and discuss the performance of our model under different conditions.

#### 5.1.1 Toy experiment 1: covariance examples

In this experiment, we compute the covariance function from a model with only one output $D=1$, three segments $Q=3$, and one latent force acting in each segment $R=1$. The switching points are defined at $t_1=1$ and $t_2=3$. For the output, we fix the decay value to be $\gamma_d=1$. We also restrict the latent forces to have the same inverse values for the length-scale parameters $\hat{\ell}_{1,0} = \hat{\ell}_{1,1} = \hat{\ell}_{1,2} = 1$, with $\ell_{r,q} = (2/\hat{\ell}_{r,q})^{1/2}$, and we fix the same values of sensitivity parameters as $S^{(0)}_{1,1} = S^{(1)}_{1,1} = S^{(2)}_{1,1} = 10$. Figure 2 shows both the resulting covariance function between the outputs, and four samples from the zero-mean GP. Dashed lines indicate the final values of each switching point. Figure 2(b) evidences that output function remains continuous across the switching points (as enforced by the definition of the model in section 3.1).

Figure 3 shows similar results than Figure 2, but one of the hyperparameters of the model is changed. In each sub-caption, we specify the hyperparameter that is modified. In each row, we show the ability of the model to perform changes in the length-scales of the forces (first row), sensitivity parameters (second row), and switching points (third row). We are interested in highlighting the flexibility of the covariance function to represent different conditions for each interval. First, if we need to infer quick time-varying outputs, it is possible to specify high values of the inverse of the length-scale parameters to deal with this requirement. Second, we can control the sensitivity parameter to specify which outputs are being driven by which latent forces (e.g. if the gene expression in a specific interval is not regulated by a particular TF, it is possible to fix the corresponding sensitivity parameter to be equal to zero). Finally, according to the last row, the covariance function has the ability to represent changes in the switching time instants.

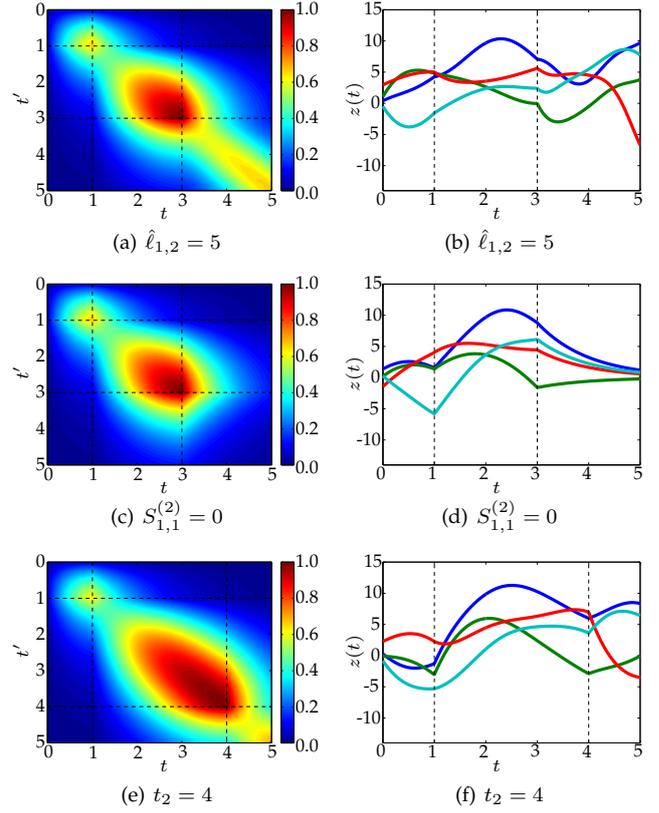

Fig. 3. Toy experiment 1. The first column shows the covariance functions between the outputs of the model, with entries given by $k_{z,z}(t,t')$. The different figures in the column are obtained by changing one parameter at the time in the covariance function from the example in Figure 2(a). The hyperparameter which is changed in each example is described in its corresponding sub-caption. The second column shows some samples generated from the corresponding zero-mean GP using the covariance functions from the first column, i.e. $y(t) \sim \mathcal{GP}(0, k_{z,z}(t,t'))$.

#### 5.1.2 Toy experiment 2: inference examples

For the following two synthetic examples, that we refer to as toy examples A and B, we generate output data by sampling from the GP with the switched LFM covariance function. We assume a zero-mean GP, i.e. $\{B_d\}_{d=1}^{D} = 0$. We sample each output for 500 data points, and add some noise with variance equal to ten percent of the variance of each sampled output. We use 200 data-points (training data) for estimating the hyperparameter of the model. The remaining 300 data-points (testing data) are used to evaluate the capacity of the model to fit the gene expression data.

**Toy example A:** we sample from a model with $D=2$, $R=1$, and $Q=3$, with switching points at $t_1=5$ and $t_2=12$. For the outputs, we fix $\gamma_1=2.0$ and $\gamma_2=1.5$. We restrict the latent forces to have the same inverse of the length-scale values $\hat{\ell}_{1,0} = \hat{\ell}_{1,1} = \hat{\ell}_{1,2} = 1 \times 10^{-3}$, but we change the values of the sensitivity parameters as $S^{(0)}_{1,1} = 10$, $S^{(1)}_{1,1} = 1$, $S^{(2)}_{1,1} = 10$, $S^{(0)}_{1,2} = 5$, $S^{(1)}_{1,2} = -10$ and $S^{(2)}_{1,2} = 1$. Note that terms $S^{(q-1)}_{r,d}$ represent the sensitivity of gene $d$ respect to protein $u_{r,q-1}(t)$ at the interval $q$. In this experiment, we want to show the ability of the model to detect changes in the sensitivities of the forces, while keeping the length-scales equal along the intervals.

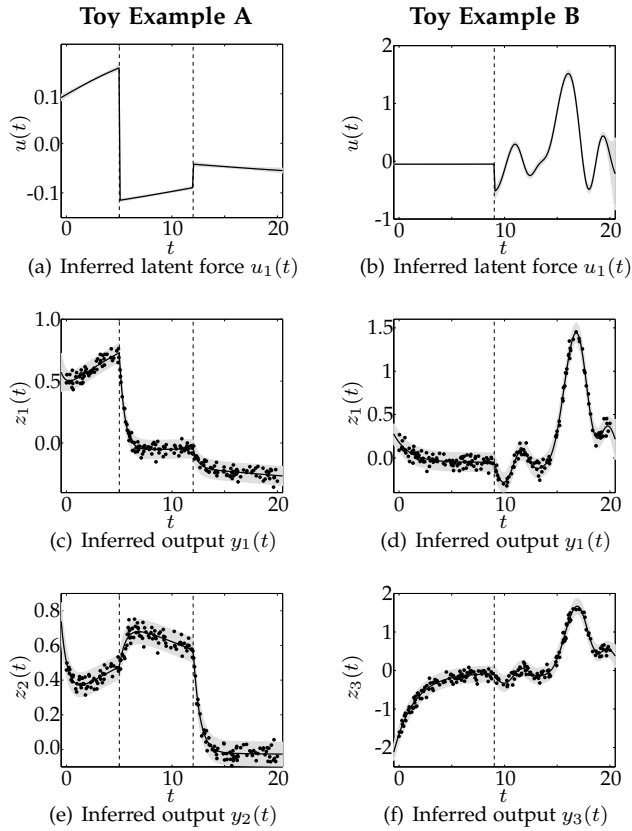

Fig. 4. Toy experiment 2. Each column shows the results obtained for both toy examples A and B from subsection 5.1.2. Prediction is performed over: the latent forces (first row), and different output profiles (second and third rows). Each figure shows the training data (dots), the mean prediction (solid lines), and two standard deviations away from the mean predictions corresponding to the grey interval. Dashed lines indicate the estimated values of the switching points.

**Toy example B:** we sample from a model with $D = 3$, $R = 1$, and $Q = 2$, with a switching point at $t_1 = 9$. For the outputs, we fix $\gamma_1 = 0.7$, $\gamma_2 = 1.5$, $\gamma_3 = 0.5$, and inverse of length-scales equal to $\hat{\ell}_{1,0} = 1 \times 10^{-3}$, and $\hat{\ell}_{1,1} = 1$. The sensitive parameters in this case are given by $S^{(0)}_{1,1} = 1$, $S^{(1)}_{1,1} = 1$, $S^{(0)}_{1,2} = 5$, $S^{(1)}_{1,2} = 1$, $S^{(0)}_{1,3} = 1$ and $S^{(1)}_{1,3} = 1$. In this experiment we want to evaluate the ability of the model to represent different oscillatory behaviours in different segments of the outputs.

Figure 4 shows the results for both toy examples A and B. The hyperparameters of the model are initialized to be equal to 1, and they are estimated (including the switching points) according to subsection 4.1. The values of the estimated hyperparameters are similar to the ones that we employed in each toy for generating the data. The inference for the toy examples A and B is shown in the first and second column, respectively. Inference results evidence that the model is able to fit either flatter (toy A) or oscillatory (toy B) output dynamics proposed in each toy example. Figures 4(a) and 4(b) also show the ability of the proposed framework to represent sudden switching changes or discontinuities in the input behaviour.

### 5.1.3 Toy experiment 3: methodology comparison

In [8], the authors proposed a toy experiment in which a TF regulator transits from an active to an inactive state. The artificial dataset is made of a single output $z(t)$ driven by a TF protein $u(t)$. The synthetic TF is defined as

$$u(t) = \begin{cases} 1, & t \in [0, 169] \cup [660, 1000] \\ 0, & t \in [170, 659] \end{cases}.$$

The parameters of the differential equation used for this example were chosen by [8] as $B = 8 \times 10^{-4}$ (basal transcription), and $\gamma = 5 \times 10^{-3}$ (decay rate). According to [8], the sensitivity parameter was fixed to be equal to $3.7 \times 10^{-3}$, which is the same during all the regimens [8]. In this experiment, we compare the results of applying the approach from [8], and of our framework. Note that the proposed TF activity presents two discontinuities at points $t = 169$ and $t = 659$. Because our proposed approach can deal with discontinuities by switching the dynamics of the input behaviour, we expect to outperform the results from [8], in both fitting the output data and estimating the artificial TF activity.

In this experiment, we fix the number of intervals $Q = 3$ at the beginning of the experiment aiming to represent the three states of the TF behaviour $u(t)$. However, we evaluated the model with different number of intervals $Q = 1, 2, 3, 4, 5$. We note that for $Q = 1$, our model corresponds to the model proposed in [7], and it infers smooth behaviours on both gene and TF inference which do not correspond to ones generated synthetically. The results are showed in [8]. We computed the log-likelihood for each model to analyse which one provides the maximum value. After the experiments, we obtained that the best model was obtained when $Q = 3$. For both approaches, the parameters $B$ and $\gamma$ were initialized to be equal to $1 \times 10^{-2}$. Due to the flat behaviour of the TF protein, the inverse of the length-scale parameters were initialized in the order of $1 \times 10^{-6}$. We initially set the sensitivity parameters to be equal to one. After the initialization, we employed the corresponding optimization modules for the estimation of the hyperparameter: Bayes expectation maximisation for the model from [8], and maximum log-likelihood for our approach. Finally, because TF concentrations have to be strictly positive, in this experiment we restricted the sensitivity parameters to be positive.[3]

Figure 5 shows the inferred output profiles (first row), and their corresponding TF activities (last row), using both inference methods. Red and blue dashed lines indicate the true output $z(t)$ and the true input behaviours $u(t)$, respectively. Figures 5(a) and 5(b) show the performance of both models to fit output data, evidencing that our approach outperforms the inference of the gene expression profile from [8]. Figures 5(c) and 5(d) show the scaled TF activity in the same order as we made for the outputs. Figure 5(c) shows that the framework from [8] exhibits a smoothed behaviour of the sudden changes in the synthetic TF activity. On the other hand, Figure 5(d) evidences how our model estimates correctly the switching instants of the latent process, outperforming the inference of the discontinuous changes from the TF activity.

---

3. This assumption does not guarantee that the latent force $u(t)$ will be strictly a positive function. However, it was enough to obtain the right results in this experiment.

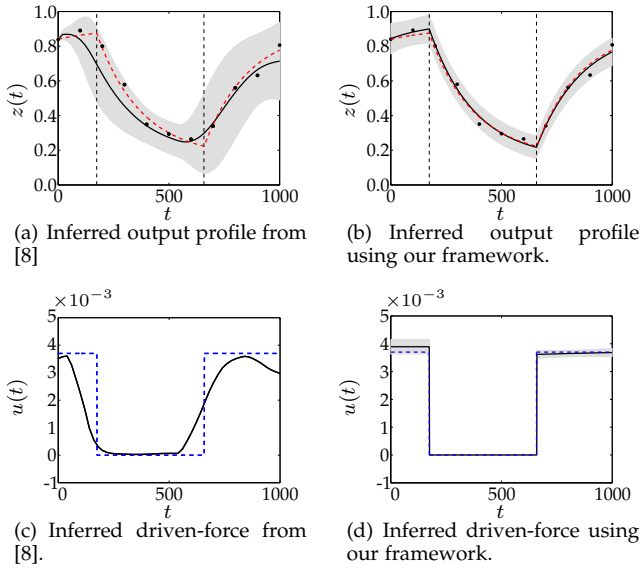

Fig. 5. Toy experiment 3. Mean and two standard deviations from the predictions over the latent force and the output in the test set. Vertical dashed lines indicate the real values of the switching points. Dots indicate training data. Red and blue dashed lines indicate the real output and TF behaviours, respectively. The reconstructed output profiles (first row), and their corresponding scaled TF activities (last row), are showed using Sanguinetti et al. [8] (first column), and employing our framework (second column). The TF activities were scaled according to their sensitivity parameters. For the inferred TF from (c), we omitted the confidence intervals in order to not clutter the figure.

Finally, our model also estimates properly the mechanistic hyperparameters. After hyperparameter estimation, our framework estimated a decay rate equal to $\gamma = 4.9 \times 10^{-3}$ against $\gamma = 4.0 \times 10^{-3}$ from [8] (true value $\gamma = 5.0 \times 10^{-3}$), and a basal transcription equal to $B = 7.2 \times 10^{-4}$ against $B = 8.0 \times 10^{-4}$ (true value $B = 8.0 \times 10^{-4}$). Respect to the inverse of the length-scale parameters, we obtained $\hat{\ell}_{1,0} = 2.2 \times 10^{-8}$, $\hat{\ell}_{1,1} = 4.8 \times 10^{-7}$, and $\hat{\ell}_{1,2} = 3.4 \times 10^{-8}$, justifying why our inferred latent function is completely flat in each segment.

## 5.2 Real data examples

For the biological examples, we train a model for each dataset according to the nature of the biological applications. Because the convergence of the model strongly depends on the initial set of the hyperparameters, they are manually initialized in each experiment in order to obtain the nearest possible approximation for fitting the gene expression data. After the initialization, the hyperparameters are optimized according to subsection 4.1.

### 5.2.1 Microaerobic shift in E. coli

In this experiment, we consider the five gene expressions from the E. coli dataset (`ompW`, `yjiD`, `hypB`, `moaA`, and `aspA`) described in subsection 4.3. We are interested in both the reconstruction of the five gene profiles and the inference of `FNR` activity. Since the activity of the regulator `FNR` is completely unknown in the dataset, we compare the results of our approach with respect to the ones obtained by [8]. For our framework, we implement a model with $D = 5$ (each output represents a different gene activity). We implement our model for different number of intervals $Q$, evaluating the log-likelihood as we did in the toy example from subsection 5.1.3. Because there are no data between 15 and 60 min, our framework for $Q > 2$ tends to over-fit the data in the first 15 min, obtaining inaccurate results. For $Q = 1$ [7], the results that we got were not satisfactory because of the different behaviours exhibited by the gene expression data before and after the instant $\sim 15$ min. Since the amount of data available in this experiment is small, predictions tended to be biologically unrealistic. For these reason, we fix a number of intervals $Q = 2$. We randomly initialize the switching point $t_1$, however we expect to obtain an estimated switching point near to 15 min where measurements were no longer taken periodically. To avoid over-parametrisation problems due to the small number of time-point available from the dataset, we tie the sensitivity parameters to be the same in each gene profile, $\{S_{1,d}^{(q)}\}_{d=1,q=0}^{D,Q-1} = \{S_{1,d}\}_{d=1}^{D}$ for any interval $q$. It means we only have one sensitivity parameter for all the regimens of each output. Then, we initialize hyperparameters with the same decay rate $\{\gamma_d\}_{d=1}^{D}$, the same sensitivity parameters $\{S_{1,d}\}_{d=1}^{D}$, and the same inverse of the length-scales $\{\hat{\ell}_{1,q}\}_{q=1}^{Q-1}$. Finally, since the measurements represent the change in concentration of mRNA relative to the initial condition, we tie the basal transcription to be equal to the decay rate, i.e. $B_d = \lambda_d$ [8].

Figure 6 and 7 show the results from [8], and the results employing the switched dynamical LFM, respectively. Figures from 6(a) to 6(e) show the reconstruction of the expression profiles using [8]. Solid lines represent the posterior mean, dotted lines represent two standard deviations for the confidence interval, and the crosses are the expression level values used as training data. Figure 6(f) shows the posterior mean for the normalized `FNR` activity using [8]. Figure 7 follows a similar structure than the one explained for Figure 6. Solid lines represent the mean prediction, region in grey represents two standard deviations for the confidence interval, and the dots are the measured expression levels used as training data. Subfigure 7(f) shows the posterior mean and two standard deviations for the normalized `FNR` activity. The dashed line indicate the estimated value for the switching point.

According to Figures 6(f) and 7(f), we note the same behaviour we detailed in the artificial experiment from subsection 5.1.3: different magnitude in the `FNR` activities. However, we observe that both profiles exhibit similar saturation behaviours. Because oxygen is the source of activation of regulator `FNR`, from Figure 7(f) it is possible to note that in the first $\sim 15$ min the bacteria starts to experiment absence of oxygen. The oxygen starvation produces a gradual growth in the concentration of regulator FNR in order to switch from an aerobic (using oxygen) to an anaerobic respiration (using different electron acceptors than oxygen). After the 15 min, we observe that the concentration of the mRNA decreases but it is different to the initial concentration, concluding that the regulator FNR remains active as long as there is no oxygen. Similar regulator dynamics were obtained in [8], [28].

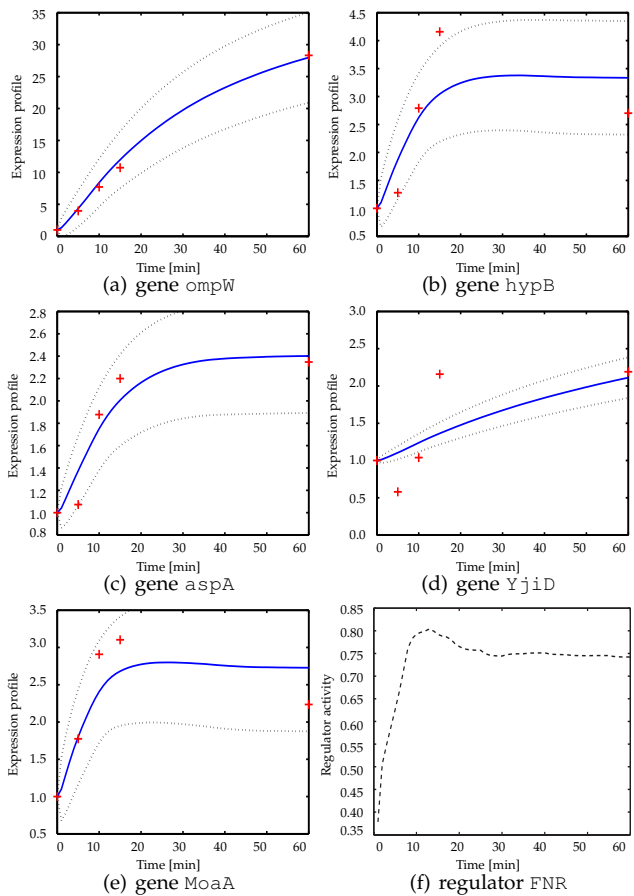

Fig. 6. Microaerobic shift in E. coli using the model proposed in [8]. Solid lines represent the mean prediction, dotted lines represent two standard deviations for the confidence interval, and the crosses are the measured expression levels used as training data. Subfigure (f) shows the posterior mean for the normalized FNR activity.

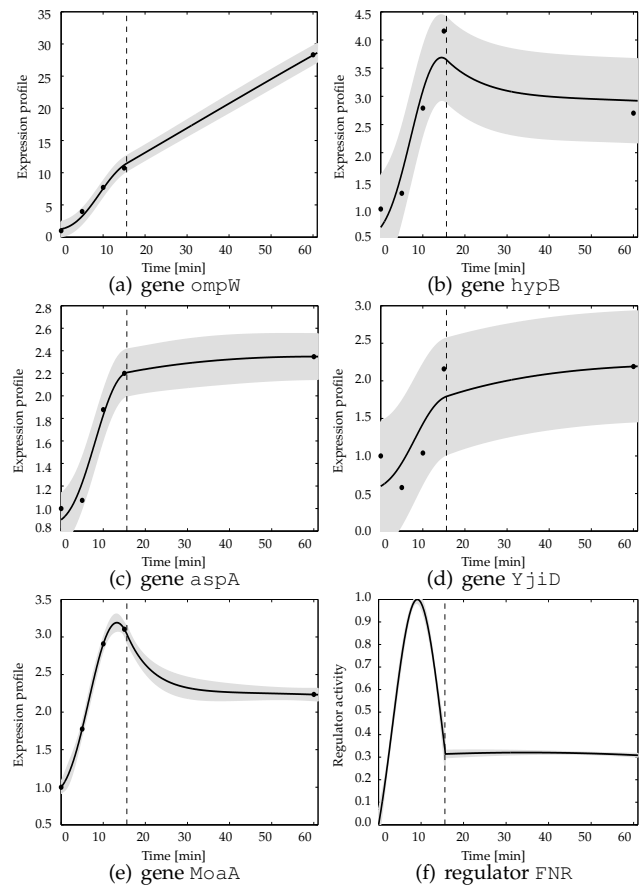

Fig. 7. Microaerobic shift in E. coli employing our framework. Solid lines represent the mean prediction, region in grey represents two standard deviations for the confidence interval, and the dots are the measured expression levels used as training data. Subfigure (f) shows the posterior mean and two standard deviations for the normalized FNR activity. Dashed line indicates the estimated value of the switching point.

From Figures 7(a) to 7(e), we observe that our framework is able to fit the expression activities of each gene. Once again, due to the switching behaviour of our model, we obtain accurate results representing changes in the gene expression data due to the changes in the FNR concentration at ∼20 min. In terms of gene expression data, our approach shows a better fit of the training data for both hypB, yjiD, and moaA profiles with reasonable uncertainties corresponding to the subfigures 6(b), 6(d) and 6(e), respectively.

### 5.2.2 Control of ribosomal production in yeast metabolism

In this experiment, we follow the same procedure as the one proposed in [18]. We consider the gene expression profiles YLR030W, TKL2, YOR359W, PFK27, RPL13A, RPL17B, and RPS16B, as outputs of our model, i.e. $D = 8$. Because the set of genes are regulated by two TF proteins, FHL1 and RAP1, we assume that two independent latent forces are driven the biological process in each interval ($R = 2$). We train a model with three intervals $Q = 3$, where each interval describes one cycle of yeast respiration. In this experiment, we also tie the sensitivity parameters per each expression profile (output) and regulator (latent force), i.e. according to the notation from the paper $S_{r,d}^{(q-1)} = S_{r,d}$ for any interval $q - 1$. For the outputs that are not regulated either by FHL1 or RAP1, we fix the corresponding sensitivity parameters to be equal to zero. We assume that the dynamical behaviour in each interval has to be similar to the other ones because they are describing the same phenomenon (one cycle of yeast respiration). In that sense, we also tie the inverse of the length-scale parameters per each latent force, i.e. $\hat{\ell}_{r,q} = \hat{\ell}_r$ for any interval $q$, in order to reduce the complexity of the model at the moment of estimating the hyperparameters.

Figure 8 shows the results of fitting the expression data. We note that our approach tends to follow the dynamics of the expression data, and that almost all the observed points are inside the confidence interval (two times the standard deviation). The results from Figure 8 are comparable to ones given by the supplemental material from [18].

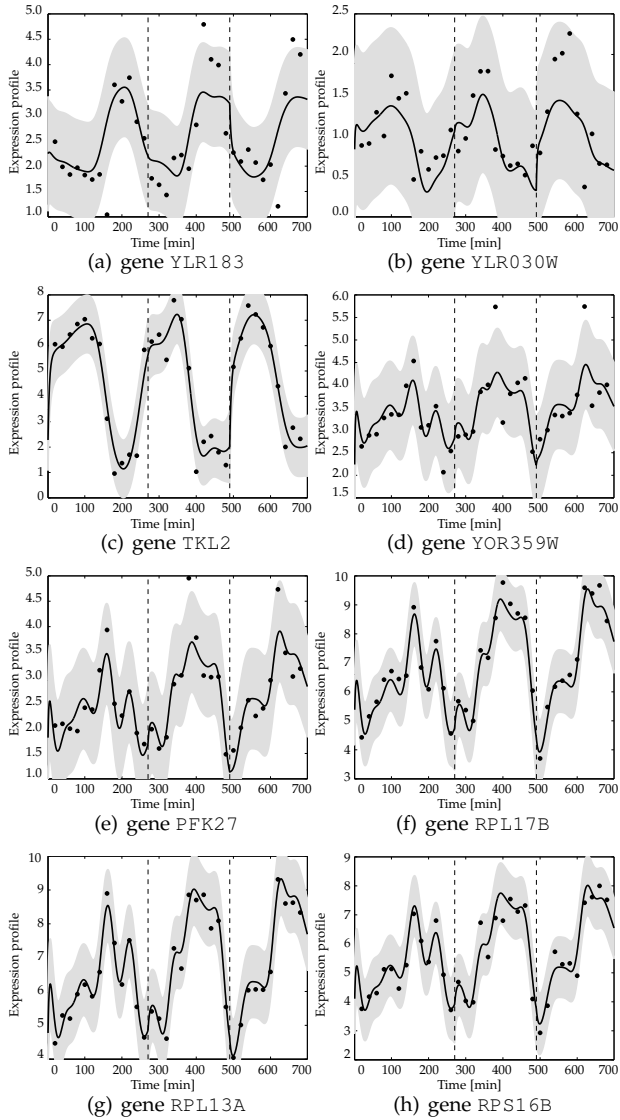

Fig. 8. Control of ribosomal protein production in yeast metabolic cycle employing our framework. Solid lines represent the mean prediction, region in grey represents two standard deviations for the confidence interval, and the dots are the measured expression levels used as training data.

With respect to the TF profiles, Figure 9 shows the inferred regulator from [18], and the inferred regulator employing the switched dynamical LFM. Figures 9(a) and 9(b) show the posterior mean profiles for both regulators FHL1 and RAP1 obtained by [18]. Figures 9(c) and 9(d) show the mean and two standard deviations for the predictions over the regulators using our proposed framework. We note that both results exhibit similar quasi-periodic behaviours in the inferred gene regulators. However, there are some differences in terms of the magnitudes for the regulator in the second interval FHL1. We believe that since each cycle refers to a similar phenomenon (backed up by the same type of behaviour observed in the actual gene expression data), it seems plausible that the behaviour of the regulator should also be similar across the cycles. Therefore, from our point of view, it seems more likely that the regulator should have the same magnitude ranges across the intervals.

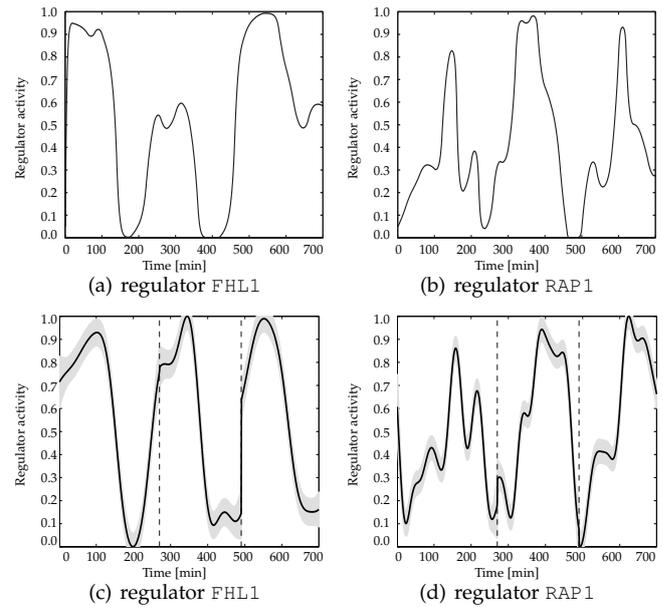

Fig. 9. Control of ribosomal protein production in yeast metabolic cycle from [18], and employing our framework. Subfigures (a) and (b) show the posterior mean profiles for both FHL1 and RAP1 regulators obtained by [8]. Using our proposed framework, subfigures (c) and (d) show the mean and two standard deviations for the predictions over the regulators. Dashed lines indicate the estimated values of switching points after the optimization. Dashed lines indicate the estimated values of the switching points after optimization. Dots indicate training data.

We can also observe from Figure 9, that the result for regulator FHL1 using [18] shows that the first and the second cycle end at ∼180 and ∼400 min (respectively), whereas that the cycles end at ∼250 and ∼500 min for the regulator RAP1. Although that each interval of our approach describes one cycle of yeast respiration for the regulator FHL1 from Figure 9(c), the estimated switching points do not necessary represent the end of each cycle for regulator RAP1. This problem is produced because our proposed model assumes that switching points for both TF profiles have to be the same, forcing the regulators to switch their corresponding transcription at the same time. The above assumption cannot be considered entirely true because the activity of the TF regulators in gene expressions commonly presents different delays for their transcription [18], [20].

Finally, according to the estimated sensitivity parameters, our framework estimates that the expressions which are being jointly regulated (RPL13A, RPL17B, and RPS16B) are more sensible to the regulator RAP1 (approximately five times greater than the contribution from FHL1). This result is consistent with the results obtained by [18], where their model predicted a minor effect of regulator FHL1 over the genes RPL13A, RPL17B, and RPS16B.

## 6 CONCLUSIONS AND FUTURE WORK

We have introduced a switched dynamical latent force model for reverse-engineering transcriptional regulation which allows the exact inference over latent transcription factors (TFs) driving some observed gene expression data through a linear ordinary differential equations. The model allows the representation of sudden changes (also discontinuities) in TF activities.

We have tested our framework in both artificial examples and real-world biological problems. According to the experimental results, the proposed approach is able to properly fit the gene expression data while also inferring TF profiles that make sense from a biological point of view. The results obtained are consistent with the previous ones obtained in the state-of-the-art. The proposed approach is also flexible to consider several assumptions a priori, e.g. to analyse the change in concentration of mRNA relative to the initial concentration, to assume different TF prior in each interval according to the biological application, to control which genes are being regulated by which TFs, etc. As we exhibited in results section, taking into account these assumptions a priori provides more accurate models that may describe the biological properties in a suitable manner.

The framework presented here could be improved in different ways. We consider that it would be interesting, for example, treating the number of intervals $Q$ and the number of latent forces $R$ as additional hyperparameters to be estimated. We would also like to generalize our approach for studying more complex biological networks, e.g. in modelling non-stationary transcriptional regulation processes where the mechanistic hyperparameters are changing in each interval, or in describing gap-gene networks dynamics where the system is driven by a non-linear combination of latent forces. Including the ability to deal with delays (as discussed in the yeast example), or to deal with large datasets in genome-wide studies, would also be some venues worth exploring.

## ACKNOWLEDGEMENTS

This work was funded by the projects "Sparse latent force models for reverse engineering of multiple transcription factors" (by Colciencias, Colombia and Universidad Tecnológica de Pereira, Colombia, under the agreement 645 of Jóvenes Investigadores); "Human-motion Synthesis through Physically-inspired Machine Learning Models" (by Universidad Tecnológica de Pereira); and "Probabilistic spatio-temporal models based on partial differential equations for the description of the regulatory dynamics for the Bicoid protein in the Drosophila Melanogaster body segmentation" (by Colciencias, Colombia and ECOS-NORD, France). MAA was Associate Professor at Universidad Tecnológica de Pereira, Colombia.

## REFERENCES

[1] U. Alon, *An Introduction to Systems Biology*. Chapman and Hall, London, 2006.

[2] G. Malacinski, *Essentials of Molecular Biology*. Jones and Bartlett Publishers, Incorporated, 2005.

[3] G. Chechik and D. Koller, "Timing properties of gene expression responses to environmental changes," *Journal of Computational Biology*, vol. 16, no. 2, pp. 279–290, 2009.

[4] L. Bintu, N. E. Buchler, H. G. Garcia, U. Gerland, T. Hwa, J. Kondev, and R. Phillips, "Transcriptional regulation by the numbers: models," *Current Opinion in Genetics & Development*, vol. 15, no. 2, pp. 116 – 124, 2005, chromosomes and expression mechanisms.

[5] L. Bintu, N. E. Buchler, H. G. Garcia, U. Gerland, T. Hwa, J. Kondev, T. Kuhlman, and R. Phillips, "Transcriptional regulation by the numbers: applications," *Current Opinion in Genetics & Development*, vol. 15, no. 2, pp. 125 – 135, 2005.

[6] L. E. Chai, S. K. Loh, S. T. Low, M. S. Mohamad, S. Deris, and Z. Zakaria, "A review on the computational approaches for gene regulatory network construction," *Computers in Biology and Medicine*, vol. 48, pp. 55 – 65, 2014.

[7] N. D. Lawrence, G. Sanguinetti, and M. Rattray, "Modelling transcriptional regulation using Gaussian processes," in *Advances in Neural Information Processing Systems 19*, B. Schölkopf, J. C. Platt, and T. Hoffman, Eds. MIT Press, 2006, pp. 785–792.

[8] G. Sanguinetti, A. Ruttor, M. Opper, and C. Archambeau, "Switching regulatory models of cellular stress response," *Bioinformatics*, vol. 25, no. 10, pp. 1280–1286, March 2009.

[9] A. Conesa, P. Madrigal, S. Tarazona, D. Gomez-Cabrero, A. Cervera, A. McPherson, M. W. Szcześniak, D. J. Gaffney, L. L. Elo, X. Zhang, and A. Mortazavi, "A survey of best practices for RNA-seq data analysis," *Genome Biology*, vol. 17, no. 1, p. 13, 2016.

[10] Z. Wang, M. Gerstein, and M. Snyder, "RNA-seq: a revolutionary tool for transcriptomics." *Nature Reviews. Genetics*, vol. 10, no. 1, pp. 57–63, Jan. 2009.

[11] G. Sanguinetti, N. D. Lawrence, and M. Rattray, "Probabilistic inference of transcription factor concentrations and gene-specific regulatory activities," *Bioinformatics*, vol. 22, no. 22, pp. 2775–2781, September 2006.

[12] M. Barenco, D. Tomescu, D. Brewer, R. Callard, J. Stark, and M. Hubank, "Ranked prediction of p53 targets using hidden variable dynamic modeling," *Genome Biology*, vol. 7, no. 3, pp. R25+, 2006.

[13] P. Gao, A. Honkela, M. Rattray, and N. D. Lawrence, "Gaussian process modelling of latent chemical species: applications to inferring transcription factor activities," *Bioinformatics*, vol. 24, no. 10, pp. 170–175, 2008.

[14] M. A. Álvarez, D. Luengo, and N. D. Lawrence, "Latent force models." in *AISTATS*, ser. JMLR Proceedings, D. A. V. Dyk and M. Welling, Eds., vol. 5. JMLR.org, 2009, pp. 9–16.

[15] ——, "Linear latent force models using Gaussian processes," *Pattern Analysis and Machine Intelligence, IEEE Transactions on*, vol. 35, no. 11, pp. 2693–2705, Nov 2013.

[16] D. J. Jenkins, B. Finkenstädt, and D. A. Rand, "A temporal switch model for estimating transcriptional activity in gene expression," *Bioinformatics*, vol. 29, no. 9, pp. 1158–1165, May 2013.

[17] M. K. Titsias, A. Honkela, N. D. Lawrence, and M. Rattray, "Identifying targets of multiple co-regulating transcription factors from expression time-series by Bayesian model comparison." *BMC Systems Biology*, vol. 6, p. 53, 2012.

[18] M. Opper and G. Sanguinetti, "Learning combinatorial transcriptional dynamics from gene expression data." *Bioinformatics*, vol. 26, no. 13, pp. 1623–1629, 2010.

[19] M. A. Álvarez, J. R. Peters, N. D. Lawrence, and B. Schölkopf, "Switched latent force models for movement segmentation," in *Advances in Neural Information Processing Systems 23*, J. Lafferty, C. Williams, J. Shawe-Taylor, R. Zemel, and A. Culotta, Eds. Curran Associates, Inc., 2010, pp. 55–63.

[20] K. Becker, E. Balsa-Canto, D. Cicin-Sain, A. Hoermann, H. Janssens, J. R. Banga, and J. Jaeger, "Reverse-engineering post-transcriptional regulation of gap genes in Drosophila melanogaster," *PLoS Comput Biol*, vol. 9, no. 10, pp. 1–16, 10 2013.

[21] J. D. Vásquez Jaramillo, M. A. Álvarez, and A. A. Orozco, "Latent force models for describing transcriptional regulation processes in the embryo development problem for the drosophila melanogaster," in *Engineering in Medicine and Biology Society (EMBC), 2014 36th Annual International Conference of the IEEE*, Aug 2014, pp. 338–341.

[22] W. Liu and M. Niranjan, "Gaussian process modelling for Bicoid mRNA regulation in spatio-temporal Bicoid profile." *Bioinformatics*, vol. 28, no. 3, pp. 366–372, 2012.

[23] P. A. Alvarado, M. A. Álvarez, G. Daza-Santacoloma, A. A. Orozco, and G. Castellanos-Domínguez, "A latent force model for describing electric propagation in deep brain stimulation: A simulation study," in *Engineering in Medicine and Biology Society (EMBC), 2014 36th Annual International Conference of the IEEE*, Aug 2014, pp. 2617–2620.

[24] C. E. Rasmussen and C. K. I. Williams, *Gaussian Processes for Machine Learning (Adaptive Computation and Machine Learning)*. The MIT Press, 2005.

[25] C. M. Bishop, *Pattern Recognition And Machine Learning (Information Science And Statistics)*.   Springer, 2007.
[26] K. P. Murphy, *Machine Learning: A Probabilistic Perspective (Adaptive Computation And Machine Learning Series)*.   The MIT Press, 2012.
[27] S. M. Kay, *Fundamentals of Statistical Signal Processing: Estimation Theory*.   Prentice-Hall PTR, 1998.
[28] J. D. Partridge, G. Sanguinetti, D. P. Dibden, R. E. Roberts, R. K. Poole, and J. Green, "Transition of Escherichia coli from aerobic to micro-aerobic conditions involves fast and slow reacting regulatory components," *The Journal of Biological Chemistry*, vol. 282, no. 15, pp. 11 230–11 237, Apr. 2007.
[29] B. Tu, A. Kudlicki, M. Rowicka, and S. McKnight, "Logic of the Yeast Metabolic Cycle: Temporal Compartmentalization of Cellular Processes," *Science*, vol. 310, no. 5751, pp. 1152–1158, Nov. 2005.
[30] C. T. Harbison, D. B. Gordon, T. I. Lee, N. J. Rinaldi, K. D. Macisaac, T. W. Danford, N. M. Hannett, J. B. Tagne, D. B. Reynolds, J. Yoo, E. G. Jennings, J. Zeitlinger, D. K. Pokholok, M. Kellis, P. A. Rolfe, K. T. Takusagawa, E. S. Lander, D. K. Gifford, E. Fraenkel, and R. A. Young, "Transcriptional regulatory code of a eukaryotic genome," *Nature*, vol. 431, no. 7004, pp. 99–104+, 2004.
[31] T. I. Lee, N. J. Rinaldi, F. Robert, D. T. Odom, Z. Bar-Joseph, G. K. Gerber, N. M. Hannett, C. T. Harbison, C. M. Thompson, I. Simon, J. Zeitlinger, E. G. Jennings, H. L. Murray, D. B. Gordon, B. Ren, J. J. Wyrick, J.-B. Tagne, T. L. Volkert, E. Fraenkel, D. K. Gifford, and R. A. Young, "Transcriptional regulatory networks in Saccharomyces cerevisiae," *Science*, vol. 298, pp. 799–804, 2002.

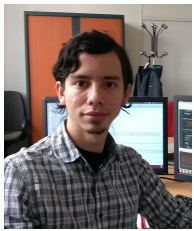
**Andrés F. López-Lopera** was born in Pereira, Colombia, in 1991. He received a degree in Electrical Engineering (B.Eng.) from Universidad Tecnológica de Pereira, Colombia in 2013, a masters degree in Electrical Engineering (M.Eng.) from Universidad Tecnológica de Pereira, Colombia in 2015. Currently, he is pursuing a Ph.D. in Applied Mathematics at the Department of Mathematics and Industrial Engineering, École des Mines de Saint-Étienne, France. His research interests include applied mathematics, machine learning, probabilistic models and signal processing.

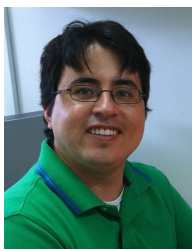
**Mauricio A. Álvarez** was born in Pereira, Colombia, in 1981. He received a degree in Electronics Engineering (B.Eng.) from Universidad Nacional de Colombia in 2004, a masters degree in Electrical Engineering (M.Eng.) from Universidad Tecnológica de Pereira, Colombia in 2006, and a Ph.D. degree in Computer Science from The University of Manchester, UK, in 2011. He was part of the academic staff at Universidad Tecnológica de Pereira, Colombia until 2016. Currently, he is a Lecturer of Machine Learning at the Department of Computer Science, The University of Sheffield, UK. His research interests include probabilistic models, kernel methods and stochastic processes.